\documentclass[11pt,letterpaper]{article}
\usepackage{amsmath,pgf,pgfarrows,pgfnodes,float,appendix, hyperref,scalerel,amssymb}
\usepackage{graphicx}
\usepackage{subfigure}
\usepackage[margin=0.9in]{geometry}
\usepackage{pgfplots}
\usepackage{tikz}
\usepackage[normalem]{ulem}
\usepackage{mathrsfs}
\usepackage{slashed}
\usepackage[compat=1.0.0]{tikz-feynman}
\usepackage{wrapfig}

\usetikzlibrary{arrows}
\pgfplotsset{compat=1.15}
\usetikzlibrary{decorations.markings}
\usepackage{braket}
\usepackage{simpler-wick}
\usepackage{mathtools}
\usepackage{rotating}
\usepackage{float}

\count\footins = 1000

\title{The holographic space-time model of inflation and its predictions for the CMB primordial spectra}
\author{Sidan A, Tom Banks\\ 
Department of Physics and NHETC\\
Rutgers University, Piscataway, NJ 08854 USA\\
\href{mailto:sa1975@physics.rutgers.edu}{sidan.aa@rutgers.edu}, \href{mailto:tibanks@ucsc.edu}{tibanks@ucsc.edu}}
\date{}

\begin{document}
\renewcommand{\figurename}{Figure}
\renewcommand\appendixpagename{\centering \large APPENDIX. }
\maketitle

\begin{abstract} 
In 1998 Carlip \cite{carlip} and Solodukhin \cite{solo} independently conjectured the equality between fluctuations in the modular Hamiltonian of black holes and its expectation value. Banks and Zurek's 2021 work \cite{BZ} generalized this observation to any causal diamonds. We use this observation to calculate the precise normalization of primordial scalar and tensor power spectra in the Holographic Space-time (HST) model of inflation. In previous work \cite{holoinflation} Banks and Fischler outlined the general calculation of inflationary power spectra in this model, which we review here.  Its conceptual basis is radically different from field theoretic inflation and it has no trans-Planckian problem, but its predictions for power spectra will be shown to fit extant data equally well, and the tensor-to-scalar ratio agrees with experimental observations \footnote{ A brief presentation can be found on the Strings 2025 website \href{https://www.youtube.com/watch?v=MDi0gzwh1zY}{[link]}.}. 
\end{abstract}

\section{Introduction}

The Holographic Space-time (HST) model of inflationary cosmology is an economical and completely finite account of the very early history of the universe.  It relies on Jacobson's \cite{ted95} view of Einstein's equations as the hydrodynamics of the BHJFSB\footnote{Bekenstein-Hawking-Jacobson-Fischler-Susskind-Bousso.} \cite{bek, hawk, ted95, fs, bousso} area law for the entropy of generic causal diamonds, to write the quantum theory of the universe as time dependent quantum mechanics in a Hilbert bundle \cite{hilbertbundles} over the space of geodesics of an FRW universe.  Along any give geodesic, at cosmic time $t$, the evolution operator is a tensor product $U_{in} (t) \otimes U_{out} (t)$.  Following Carlip \cite{carlip} and Solodukhin \cite{solo}, as generalized in \cite{BZ, hilbertbundles, tbwflm}, $U_{in} (t)$ is a time ordered product of unitary embedding operators of the form $e^{- i L_0 (t_i)}$, where the discrete times are separated by one Planck unit.  At each time $L_0$ is the Virasoro generator of an abelian Thirring model \cite{hilbertbundles} built from two-dimensional fermion fields labeled by angular momentum multiplets on the two sphere (the holographic screen of the cosmological apparent horizon).  One angular momentum multiplet is added with each Planck time step.  The fields live on an interval (the stretched apparent horizon) and have a UV cutoff which is determined by the smallest $t$ for which the Carlip-Solodukhin hydrodynamic analysis of the dynamics is applicable.  In principle, this can only be determined by experiment/observation.  Theoretically, there must be enough $1 + 1$ dimensional momentum modes for Cardy's formula to capture the spectral density.  This model reproduces the Einstein equation of state of the flat FRW universe with scale factor
\begin{equation}
    e^{a(t)} = \sinh^{1/3} (3t/R_I) ,
\end{equation}
if we stop increasing the size of the Hilbert space at $t \sim R$ and allow time to evolve with the final $L_0 (R_I)$ as Hamiltonian.  This universe begins as a finite quantum system, evolves quickly to a sequence of cutoff $1 + 1$ dimensional conformal field theories, whose hydrodynamics is that of the flat $p=\rho$ universe, and asymptotes to a de Sitter universe with radius $R_I$ \cite{tbwflm}.

For a general space-time, $U_{out} (t)$ is determined by the Quantum Principle of Relativity (QPR): {\it For any pair of intervals along two geodesics, there is a maximal area causal diamond in the overlap of the causal diamonds corresponding to those intervals.  That diamond must be identified with isometric tensor factors in each of the fiber Hilbert spaces, and the density matrices assigned to those factors for every choice of initial conditions must have equal entanglement spectra.}  For the above space-time one can satisfy the QPR for any choice of $U_{out}$ by choosing the tensor factor in the Hilbert space of two trajectories whose overlap diamond at time $t$ has area $A_O$ to be the ``in" Hilbert space of each at time $\sqrt{A_O}$.   The dynamics of our system is built so that for generic initial states along any trajectory, the modular Hamiltonian at any time is the $L_0$ generator of the same Thirring model, with the number of fermions determined by the area.  So the QPR is automatically satisfied.  The QPR puts constraints on $U_{out}$ when we discuss space-times with constrained initial states.  In that case a constraint on the past horizon of some causal diamond must have an antecedent on the past horizon of a larger diamond that contains it, and the dynamics on that larger diamond must be consistent with that on the smaller diamond it contains.   In a moment we will discuss more realistic cosmologies, where such issues arise.  However, we will cheat by using the semi-classical language of particle physics, where the QPR reduces to the usual rules of space-time general covariance.  Deriving those rules from the quantum model is a task that remains to be accomplished.

The general formalism on which this paper is based was developed in a series of papers going back $25$ years.  We cannot go through all the details of that formalism here without making the current work too long and indigestible.  The relevant references can be found in \cite{keyholo}.  In Appendix \ref{app:review} we give a brief review of how a model of matrix valued fermion quantum fields $\psi_i^A (z,t)$ can produce qualitative properties of black hole physics in de Sitter space. 

The model described above will be used as a component of a more realistic model of an Inflationary universe.  It describes one horizon volume of inflation before the slow-roll era starts, in a completely non-singular quantum mechanical way.  It saturates the covariant entropy bound at every time and has no localized excitations and it is never well approximated by quantum field theory in curved space-time.  Independent horizon volumes are considered separate physical systems, initially out of causal contact, rather than arising from short wavelength modes in a single horizon volume which get stretched to larger size.  This is the way HST models avoid the trans-Planckian problems of field theory inflation.  

Finally we should note that HST models are conceptually very different from other models of ``holographic cosmology" that have appeared in the literature \cite{otherholo1, otherholo2}.  They are characterized by having all fundamental degrees of freedom localized on the boundaries of causal diamonds in a background space-time.  In the current version of the models, these degrees of freedom form a $1 + 1$ dimensional CFT built from fermion fields labeled by solutions of the Dirac equation on the holographic screen of a diamond.  Bulk excitations in the diamond are realized as constrained low entropy states of these variables.

\section{From Holographic Inflation to the Hot Big Bang}

As stated, the model described above has no localized excitations.  The Schwarzschild de Sitter (dS) black hole entropy formula shows that localized excitations in the asymptotic de Sitter space in which we apparently live are constrained states of a finite dimensional quantum system \cite{tbwfdS}.   The static dS energy is related to the number of constrained q-bits by $2\pi R E = N {\rm ln}\ 2$.  An explicit quantum model of this was first constructed in \cite{bfm}.  For a fixed total energy, the maximal localized entropy concentrates all of the energy in a single black hole.  

We've seen that a universe that initially saturates the covariant entropy bound rapidly approaches empty dS space if it has a finite number of states.   In order to avoid this, we must build a model which has a constrained initial state.  That is, after some period of following the maximal entropy model, we continue to let the ``in" Hilbert space describing the cosmological causal diamond expand, but constrain the states so that the subsystems describing different copies of the dS horizon volume no longer interact.  In the matrix model language of \cite{bfm} (enhanced by promoting the fermionic oscillators to cutoff two-dimensional fermion fields as in \cite{hilbertbundles}) we set initial conditions so that off diagonal components between different blocks of matrices bilinear in the fermions vanish in the initial state.  The time scale for interaction between the blocks, in the proper time along any geodesic, is the inverse of the proper distance to the apparent horizon.  

It is extremely important at this point to remember the lesson of \cite{ted95}: solutions of Einstein's equations describe hydrodynamics of quantum systems.  In a quantum system we are free to choose the time dependent Hamiltonian any way we like.  Thus, the slow-roll metric which expands the physical horizon corresponds to a choice of the rate at which we transfer degrees of freedom from $U_{out}$ to $U_{in}$.  In geometric language, this is a choice of how fast the physical size of the cosmological diamond expands. We can represent it in Einstein's equations as an effective pressure and energy density if we insist on doing this in a homogeneous and isotropic manner.  This is a choice of model, not a fine tuning of initial conditions. It means we choose the same $U(t)$ for every fiber of our Hilbert bundle and we insist that all of the Hilbert spaces carry representations of $SU(2)$.  The general construction of \cite{hilbertbundles} has a Hamiltonian on each diamond boundary that is formally invariant under ``fuzzy area preserving diffeomorphisms", but we are requiring something slightly more than that.  We require an identification of $SU(2)$ subgroups of each of those large groups, such that the new degrees of freedom added in each Planck time step consist of exactly one angular momentum multiplet of the same $SU(2)$.  This is the way isotropy is enforced on the model.  We will discuss homogeneity and isotropy of the {\it state} of the system below.  

Note that as long as $p + \rho \geq 0$, we can always represent an FRW universe as sourced by a scalar ``inflaton" field, but that field has no quantum dynamical significance in the HST model.  It is just a bookkeeping device, which we will use only to make contact with equations in the existing literature on inflation.   As a consequence, none of the constraints on inflation that are supposed to follow from string theory or the weak gravity conjecture apply to the HST model.  The only constraint that we have been able to find is that during slow-roll, the universe must expand sufficiently fast that there is not enough time to equilibrate the off diagonal matrix elements and combine that individual horizon volumes into a single equilibrated system.  Roughly speaking, this constraint is
\begin{equation}
    \epsilon \equiv - \frac{\dot{H}}{H^2} > C ({\rm ln} {M_P/H})^{-1} , 
\end{equation}
Since we also have, from the validity of semi-classical space-time pictures $H \ll M_P$ and (as we'll see) from the fit to the Cosmic Microwave Background (CMB) 
\begin{equation}
    \frac{H}{\epsilon} \sim 10^{-4} ,
\end{equation}
we conclude that, for $C$ of order $1$, $\epsilon \sim 0.1$. We will see in section \ref{sec:angmomratio} that given the ratio between tensor and scalar perturbations in HST models, this is perfectly compatible with observation.

The basic claim is that, to a detector looking back at the history of the universe starting from the end of the slow-roll era, the universe is homogeneous and isotropic except that there is a distribution of localized finite entropy quantum mechanical systems, which have identical modular Hamiltonians.  Bekenstein and Hawking tell us that we should interpret these as black holes, and that the modular fluctuations are fluctuations in the black hole mass.   The Kerr black hole entropy formula then tells us to expect angular momentum fluctuations as well.  We interpret these as processes in which the black holes emit and absorb stochastic gravitational waves.  We emphasize that the Carlip-Soldukhin formula tells us only about entropy fluctuations, which are the same as mass fluctuations.  For a fixed mass, the Kerr formula tells us that angular momentum zero is favored and allows us to compute the fluctuations around that central value. Our fit to the density fluctuations predicts that there is on average one black hole per horizon volume at the end of the slow-roll era, so we compare the Kerr prediction for angular momentum fluctuation in a horizon volume to that predicted by classical cosmological perturbation theory, in order to normalize the two point function of the latter.

The remainder of this paper will be devoted to calculating the spectra of primordial density fluctuations expected and determining the tensor-to-scalar ratio from this picture. The computation of the spectra of the primordial fluctuations was first done by \cite{Mukhanov:1981xt, Starobinsky:1982ee, Hawking:1982cz, Guth:1982ec, Mukhanov:1982nu, Bardeen:1983qw}, and a review on the theory of cosmological perturbations is given by \cite{Mukhanov:1990me}. We borrow equations from the masterful paper of Maldacena \cite{maldafluct}, which gives a concise outline of the classical theory of cosmological fluctuations in co-moving gauge. The HST model provides the appropriate initial conditions for the classical solutions. Two other points should be noted before we begin. The black holes of HST models decay, giving rise to the Hot Big Bang, and their decay produces an additional spectrum of stochastic gravitational waves, with a different spectral shape. The size of this contribution depends on the number of particle species that are effectively relativistic at the reheat temperature.  Since that temperature is $\sim 10^{10}$ GeV in HST models, we only have a lower bound on that number. It is likely that one of these two components is dominant, but both are small and it is not clear which is the dominant one. In this paper we will assume that the gravitons from black hole decay are negligible. In addition, density fluctuations grow during this early matter dominated era and go non-linear at a time parametrically shorter than the black hole decay time. Thus, bound structures will be formed.  If some of the black holes carry a discrete $Z_N$ quantum number and leave over stable remnants, they can be primordial black hole dark matter. Many of the decay products of the inflationary black holes will be quite massive and may be captured into early structures centered around those stable remnants.  We will not discuss the implications of this for the formation of the earliest galaxies.  

\section{Classical Fluctuations and Their Initial Conditions}

\subsection{The HST Framework}\label{sec:HSTframework}

The basic point of view of the HST models is that of Jacobson \cite{ted95}: the classical Einstein equations are the hydrodynamic equations of quantum gravity, as encoded in the area law for the expectation value of the modular Hamiltonian of a causal diamond.  The classical cosmological fluctuation equations make perfect sense in this context.  

The second key ingredient, as we have outlined above, is that the most probable cosmological initial condition containing localized excitations begins as multiple isolated copies of the flat $p = \pm \rho$ cosmology, with some fixed dS radius $R_I$\footnote{The universe begins as a finite quantum system, evolves quickly to a sequence of cutoff $1 + 1$ dimensional CFTs, whose hydrodynamics is that of the flat $p=\rho$ universe, and asymptotes to a de Sitter universe with radius $R_I$ \cite{tbwflm}.}, which then manifest as a distribution of black holes in an expanding universe whose horizon grows larger than $R_I$.  The black holes are finite quantum systems and will give rise to unavoidable fluctuations in entropy (black hole mass) and angular momentum, which we will compute in this paper.  In addition, there might be inhomogeneities or anisotropies in the distribution of black holes in some set of background FRW coordinates.  

This possibility can be ruled out by both arguments based on our specific models and on very loosely defined {\it bio-trophic} grounds.  Our models treat the black holes as equal size diagonal blocks in a matrix with single trace dynamics (see Appendix \ref{app:review} and the references), and inherit an $S_N$ gauge symmetry between the individual blocks from the unitary symmetry of the underlying model.  The homogeneity and isotropy of the probability distribution in the hydrodynamic approximation of the state of the system follows from this gauge symmetry.

From a more phenomenological point of view, the principle that one should start with the most probable initial conditions allowing for localized excitations implies that the initial density of the black hole gas is such that they barely escape merging back into a uniform $p = \rho$ universe. Inhomogeneities in the probability distribution of black holes would quickly lead to mergers and the formation of large black holes in which all of the matter in the universe would be trapped. This would lead to a universe very unlike our own, dominated by large black holes. The small inhomogeneities coming from quantum fluctuations allow enough time \cite{tbas} for most of the black holes to decay into ordinary matter.  Banks and Fischler \cite{tbwfgravessay} have recently speculated that some of the structures formed during this early black hole dominated era could evolve into the seeds for the supermassive black holes which appear to exist in even the very earliest galaxies.  We will not discuss this possibility in the current paper.  

Curvature and the possibility of toroidal compactification of flat cosmologies are dealt with in a manner similar to the maximal entropy situation.  Negative curvature leads to a universe which is never close to saturating the covariant entropy bound, while positive curvature or tori lead to re-collapse on what are generically microscopic time scales.  Recall that the maximal entropy universe predicted flat spatial sections uniquely, simply from its quantum dynamics, with no fine tuning of initial conditions.

The upshot is that we should assume that the probability distribution of black holes is homogeneous and isotropic in flat FRW coordinates.  As outlined in \cite{keyholo} it has an additional approximate symmetry, whose origin is the individual $L_0$ generators of the quantum systems on each black hole/inflationary horizon.  Locally, these act like the dS time translation operator, and rescale momenta that are of order $R_I^{-1}$ in the vicinity of each horizon.  The sum of all of these generators acts on the probability distribution of black holes and approximately satisfies the sub-algebra of $SO(1,4)$ which preserves a flat coordinate patch of dS space.  This is enough to prove approximate dS invariance of the two-point functions of the black hole probability distribution.  This implies dS invariance of the tensor power spectrum, but the gauge invariant formula for the scalar fluctuations differs by a factor of $\epsilon^{-1}$, where $\epsilon = - \frac{\dot{H}}{H^2}$, from the fluctuation in black hole entropy, so this spectrum will show a deviation from the dS invariant prediction.  It's ironic that this inflationary model has (part of) the dS isometry group as a global symmetry while it has long been argued \cite{tbandprevious} that the same would not be true of the quantum theory of a space-time that was exactly dS space.

Let us explain the origin of the different powers of $\epsilon$ in our fluctuation formula.  In field theory inflation models the gravitational field is quantized and the origin of CMB fluctuations are the quantum fluctuations of the scalar mode of this field in co-moving gauge.  The Einstein-Hilbert action provides a $\sqrt{\epsilon}$ in the normalization of this field, and predicts that the fluctuation $\frac{\delta H}{H} \sim \sqrt{\epsilon}$.  By contrast, in our model $\frac{\delta H}{H}$ is independent of $\epsilon$ and comes from the Carlip-Solodukhin formula for the fluctuation of black hole entropy.  The {\it relative} fluctuation of angular momentum per black hole is then derived from the Kerr black hole entropy formula.  

The other major difference between our model and field theory models is that field theory models calculate correlation functions of quantum fields in the background slow-roll geometry and plug those into the formulae for classical cosmological perturbation theory.  In our model, up to the factor of $\epsilon^{-2}$ explained above, in the gauge invariant scalar correlator, we view the two-point functions of cosmological perturbation theory is reflecting approximately de Sitter invariant probability distributions for black holes and fit the normalization of those two-point functions to the Carlip-Solodukhin and Kerr formulas.  

Given that the CMB data is a single function of angular momentum plus a bound on the as yet unseen B modes, it is no surprise that we can fit the data with our free function $\epsilon (t)$.  Indeed, the same can be said of any model of inflation.  The thing that is relevant is how close the data is to scale invariance.  In our case things are even more surprising and we found that the data was just barely compatible with all of the theoretical constraints on our model.  

The meaning of the phrase ``horizon crossing" for a perturbation of a particular wave number is somewhat different in HST models than in conventional field theory.  The basic premise of HST inflation models is that for causal diamonds smaller than the inflationary radius $R_I$, the description of quantum gravity by quantum field theory or any other theory of bulk local excitations is incorrect.  The proper quantum theory consists of a finite set of q-bits localized on the holographic screens of nested causal diamonds interacting as a cut-off $1 + 1$ dimensional field theory of matrix valued fermion fields. The dynamics is time dependent and the size of the matrices increases with FRW time, in a way that mirrors the expansion of the cosmological horizon \cite{hilbertbundles,tbwflm}.  

Local bulk physics begins to be relevant at length scales $>R_I$ because of an {\it assumption} of constrained initial conditions. In terms of the fundamental matrix model, these are conditions that set the initial state so that the matrix fields bilinear in fermions are block diagonal, with blocks of size $(R_I/L_P) \times (R_I/L_P)$, for the period of time conventionally called ``slow-roll inflation".  The length of this period is determined, for theorists, by the choice of initial conditions and the size of the full Hilbert space of the system.  Phenomenologically it is determined by the requirement that inflation last long enough to expand wavelengths of order $R_I$ to the size of the current horizon.  Before slow-roll began, the hydrodynamic Einstein equations had only homogeneous isotropic solutions consistent with the quantum dynamics. After slow-roll begins we have isolated systems and we can imagine fluctuations that transfer physical quantities between them.  This is the point where local bulk physics, at least in the hydrodynamic sense, begins to make sense.  We've argued \cite{keyholo} that that physics should still be approximately homogeneous and isotropic, with the fluctuations arising primordially from quantum fluctuations of the individual isolated systems.  However, if we work in wave number space, fluctuations of $|k| < R_I^{-1}$ manifest as statistical fluctuations of the hydrodynamics variables, which, following Jacobson \cite{ted95} we identify with the gravitational field.  So a mode ``crossing the horizon" means the point at which its wavelength is long enough for it to be considered a statistical fluctuation in the {\it spatial distribution} of isolated systems, rather than a quantum fluctuation of any particular system.  

From the bulk hydrodynamic point of view, our constrained initial conditions correspond to initial conditions on a flat FRW slice, in which we have a distribution of black holes, sufficiently well separated and homogeneous that they do not begin to merge until fluctuations in the distribution become of order one.  As we'll see, the explicit calculations show that there is approximately one black hole per cosmological horizon volume at the end of slow-roll inflation.  These black holes should be thought of, during the entire early universe, up to the end of the slow-roll era, as what are conventionally called inflationary horizon volumes.  The identification of black holes with horizon volumes has been the key to the HST resolution of the trans-Planckian problems of field theory inflation models since \cite{keyholo}.  The conjecture of a universal fluctuation formula for the modular Hamiltonian of causal diamonds reinforces this conclusion.  

While the use of classical general relativity as a local hydrodynamic approximation to the quantum theory is valid during the slow-roll era, there is no detailed picture of the system in terms of local excitations in the bulk. We will see that the physical size of the horizon\footnote{The physical size of the forward light cone.} expands only a bit during the slow-roll era, so an intuitive local picture of each horizon volume is that the horizon expands, with a large black hole of near horizon size in it.   A more conventional local picture becomes relevant beginning on the FRW surface at which inflation ends.  As indicated above, we define this by the number of e-folds required to stretch a co-moving length $R_I$ out to the current cosmological horizon. This choice is influenced by our belief that we live in an asymptotically dS universe with cosmological horizon close to the current size of our causal diamond.  One could always imagine, as many field theorists do, inflation that goes on much longer. However, in models based on the principles of HST, which have strict causality built into their mathematical structure, future evolution away from that hypothetical asymptote could only affect observations that have not yet been done.  

On that first post-inflationary surface we invoke the principle, based on the equalities of entropies, sizes, and fluctuations of the modular Hamiltonian\cite{carlip,solo,BZ,tbpddS}, that the isolated inflationary horizon volumes manifest as a collection of identical black holes, traveling on geodesics in the FRW metric. The black holes are dispersed in space with a probability distribution whose cumulants are calculated by cosmological perturbation theory during the slow-roll era.  However, since the surfaces of constant black hole density are obviously the co-moving gauge surfaces, it is the fluctuations $\frac{\delta H}{H}$ rather than the gauge invariant field $\zeta$, which should be compared to predictions from the Carlip-Solodukhin ansatz. After this point we can evolve the black holes using classical GR/Newtonian dynamics, supplemented by Hawking decay formulae.  

To make the comparison between cosmological perturbation theory and the HST model, we equate the integral of the correlation function of $\frac{\delta H}{H}$ over the horizon volume at the end of inflation, to the number density of black holes per volume times the Carlip-Solodukhin fluctuation per black hole.  While the power spectrum depends on the constant $b$ normalizing the slow-roll factor $\epsilon$ and the overall normalization $N$ of the $\frac{\delta H}{H}$ correlator, only through the combination $b^{-2} N$, the Carlip-Solodukhin ansatz fixes $N$ in terms of the density $n_{BH}$ at the end of inflation. The formula for $n_{BH}$ is
\begin{equation}
    n_{BH} =\frac{1}{V_{end}},
\end{equation}  
where $V_{end}$ is the horizon volume at the end of inflation, because the physical size of the horizon does not expand very much during slow-roll.  

We also note that in the HST model, the ratio of tensor to scalar fluctuations is determined by using the Kerr black hole entropy formula to estimate the angular momentum fluctuation per black hole.  We can then calculate the angular momentum per black hole from the cosmological perturbation formulae by simply calculating $\langle J^2 \rangle$ from the tensor two-point function by Wick's theorem and dividing by the total number of black holes.  The calculation then fixes the tensor to scalar ratio $r$ in terms of the parameter $b$, because $n_{BH}$ drops out of the ratio.  We find in particular that $r$ scales like $\epsilon^2$ rather $\epsilon$ because the HST model does not have the $\sqrt{\epsilon}$ suppression of scalar fluctuations that one finds in field theory models.  The latter follows from the canonical normalization of the scalar mode of the gravitational field in the Einstein Hilbert Lagrangian and the assumption that scalar fluctuations are quantum fluctuations of this field.  In fact, since the entropy bound is saturated in an inflationary horizon volume, field theory is not a good approximation to most of the degrees of freedom there \cite{CKN}. 

The additional suppression of tensor fluctuations is important because of a theoretical lower bound on the rate of slow-roll expansion.  At the level of a fundamental quantum model of matrix valued fields, the isolation of individual black holes/inflationary horizons is a consequence of an initial condition setting off diagonal matrix elements of matrices bilinear in the fields to zero.  The Hamiltonian is a single trace of the matrix fields. This is a condition on all of the degrees of freedom that will ever be encountered in the universe.  The dynamics enforces causality by having a time dependent Hamiltonian, which only couples together degrees of freedom belonging to a given causal diamond.  The natural time scale for interactions of boundary degrees of freedom in a diamond is the total proper time in the diamond\footnote{In this way the background classical geometry acts as a hydrodynamic guide to the construction of the quantum theory.}.   Trace dynamics of this type is almost certainly a fast scrambler \cite{ss}.  Thus, if we do not constantly increase the number of vanishing off diagonal matrix elements, the system will rapidly eliminate the initial constraints and maximize its entropy.

Since we are dealing with time dependent Hamiltonians, we are completely free to increase the rate at which we add degrees of freedom (and off diagonal constraints on initial conditions) to make sure that the black holes remain isolated.  From the space-time point of view this puts a lower bound on the expansion rate of the universe.  It is 
\begin{equation} \epsilon > C ({\rm ln}\ (M_P/H))^{-1}.\end{equation}
The constant $C$ is hard to calculate.  We assume it is of order one.  The argument of the logarithm is the number of constrained q-bits that are added when the horizon radius\footnote{The size of the causal diamond at this time, i.e. the cosmological apparent horizon.} is of order $H^{-1}$. During slow-roll, $H$ is close to $R_I^{-1}$ and fits to the CMB fluctuation data will give us $(\epsilon R_I M_P)^{-1} \sim 10^{-4}$.  Since slow-roll means $\epsilon < 1$, the combination of these conditions implies that $\epsilon \sim 0.1$.  This would be inconsistent with the conventional formula for the tensor-to-scalar ratio, but it is consistent with the HST formula.  

Since the fit to scalar CMB data alone gives us only a constraint on $b^{-2}A$, we cannot fix all of the parameters of the model from data. We can however, check for consistency.  In particular, we find that keeping $\epsilon $ less than $1$ and not smaller than the fast scrambling bound is consistent only because the scalar spectral index is so close to $1$.  $\epsilon$ varies exponentially with proper time (in units of $R_I$) along FRW geodesics but the coefficient in the exponent is proportional to $n_s - 1$ so it is relatively unchanged over many e-folds.  This also implies that, at the end of inflation, the horizon has not expanded very much and the black hole number density is order one in units of $R_I^{-3}$.  That is, after the end of slow-roll inflation we have a dilute gas of black holes with one black hole per horizon volume, with the black hole radius only somewhat smaller than the horizon radius.  It turns out that our fit is consistent with our theoretical bound on $\epsilon$ only because $n_s$ is so close to $1$.  This also fixes $ b \sim 0.1$, so that despite the fact that the equations are under determined, we are able to fit all the parameters in our model. Data fixes the coefficient $A$ and matching to the Carlip-Solodukhin ansatz fixes the inflationary radius $R_I$.  

We would be remiss if we did not mention another major assumption we have made in our discussion of tensor fluctuations.  At FRW times later than those considered in this paper, most of the black holes will decay and their decay products will include gravitons.  Those gravitons will contribute to the tensor modes of the CMB.  Their spatial distribution mirrors that of the scalar fluctuations, rather than the dS invariant distribution of tensor modes we've calculated here.  The probability of graviton production is inversely proportional to the number of effectively massless species produced at the Hawking temperature of these very tiny black holes.  This is impossible to estimate given our present knowledge of particle physics. We have assumed that this contribution to the tensor spectrum is negligible compared to the one we have calculated.  

\subsection{Primordial power spectrum of the scalar perturbations}
Start with the Lagrangian for gravity and a massless scalar field $\phi=\phi(t)$, and set $\hbar = c = G = M_P^{-2} = 1$ in the Einstein-Hilbert action, 
\begin{equation}
    \begin{split}
        S = \int d^4x \sqrt{-g} \left [\frac{R}{16\pi} - \frac{1}{2} g^{\mu\nu}\partial_{\mu}\phi\partial_{\nu}\phi - V(\phi) \right].
    \end{split}
\end{equation}
We emphasize that in single field inflation, a scalar field is just shorthand for a mild inequality on a general FRW metric.  Indeed, given an FRW metric, we can use Einstein's equations to define the pressure and energy density in terms of the FRW curvature tensor.  As long as $p + \rho$ is non-negative we can fit these to a scalar field with a potential. Another indication of this is the existence of co-moving gauge for fluctuations:  the fluctuations are all fluctuations in components of the metric tensor.  We will use the conventional notation in order to borrow familiar equations from the famous paper of Maldacena \cite{maldafluct}. A homogeneous solution, representing an expanding universe with flat spatial slices, has the metric,
\begin{equation}
    ds^2=-dt^2+e^{2a(t)}dx_idx_i = e^{2a(t)}(-d\tau^2+dx_idx_i),
\end{equation}
where $d\tau=e^{-a(t)}dt$. The slow-roll parameters are 
\begin{equation}
    \begin{split}
        \epsilon &= 
        \frac{1}{2}\frac{\dot{\phi}^2}{\dot{a}^2}, \hspace{10mm} \text{and} \hspace{10mm} 
        \eta = - \frac{\ddot{\phi}}{\dot{a}\dot{\phi}}. \\
    \end{split}
\end{equation}
In comoving gauge
\begin{equation}
    \begin{split}
        \delta \phi &= 0 , \hspace{10mm} h_{ij} = e^{2a}[(1+2\zeta)\delta_{ij}+\gamma_{ij}] \hspace{5mm}
        \text{with} \hspace{5mm} \partial_i\gamma_{ij}=0 \hspace{5mm} \text{and} \hspace{5mm} \gamma_{ii}=0,
    \end{split}
\end{equation}
where $\zeta(\vec{x},t)$ is the fluctuation in the scalar field, and the fluctuation in energy density $\rho$ is $\delta \rho \propto \delta V(\phi) = V'(\phi_0)\zeta(\vec{x},t)$. These fluctuations can be observed in the CMB as temperature fluctuations. Focusing on scalar modes, we write down the action for scalar fluctuations in momentum space,
\begin{equation}
    \begin{split}
        \zeta(\tau, \vec{x}) &= \int \frac{d^3k}{(2\pi)^3} \hspace{1mm} \zeta_{\vec{k}}(\tau) e^{i\vec{k}\cdot \vec{x}},  \\
        S_{scalar} &= \int \frac{dt d^3k}{(2\pi)^3} e^{3a(t)}\epsilon(t)\big[\dot{\zeta}^2-e^{-2a(t)}k^2\zeta^2\big].
    \end{split}
\end{equation}
Perform a change of variable and write the above expression as an integral over $a(t)$,
\begin{equation}\label{eq:ateta1}
    \begin{split}
        da &= \dot{a} dt = Hdt, \hspace{10mm} \epsilon = -\frac{\dot{H}}{H^2} = -\frac{H_a}{H},
    \end{split}
\end{equation}
where $H_a = dH/da$. Then the action becomes
\begin{equation}\label{eq:zetaeom}
    \begin{split}
        S_{scalar} &= -\int \frac{da d^3k}{(2\pi)^3}  \hspace{1mm} \left(e^{3a} H_a \zeta_a^2 - k^2 e^a H_a H^{-2}\zeta^2 \right),
    \end{split}
\end{equation}
and the equation of motion is given by  
\begin{equation}
        \zeta_{aa} + \left(3+H_{aa}H_a^{-1}\right)\zeta_a + k^2H^{-2}e^{-2a}\zeta = 0.
\end{equation}
Let $D = 3+\sigma$ where $\sigma = H_{aa}H_a^{-1}$. Treat $H_{aa}H_a^{-1} \ll 1$ as a constant, and change variable with $Y=kH^{-1}e^{-a}$, then the equation of motion becomes
\begin{equation}
    \begin{split}
        Y^2\zeta_{\vec{k},YY} -(D-1) Y\zeta_{\vec{k},Y} + Y^2\zeta_{\vec{k}} = 0.
    \end{split}
\end{equation}
The solution is given by
\begin{equation}
    \begin{split}
        \zeta_{\vec{k}}(\tau) &= Y^{\frac{D}{2}}\left(c_1 J_{\frac{D}{2}}(Y) + c_2 Y_{\frac{D}{2}}(Y)\right).
    \end{split}
\end{equation}
$J_n(z), Y_n(z)$ are Bessel functions of the first and second kind, respectively. The constants $c_1, c_2$ are determined by de Sitter symmetries and the requirement that the only singularity to be when two points coincide \cite{quantumfieldsincurvedspace}. the scaling symmetry should be invariant under translation and rotation, and in the limit $k, \tau \rightarrow \infty$ it must reduce to $e^{-ik\tau}/\sqrt{2k}$. This implies that $c_2 = ic_1$ and $c_1 = 1/(Y\sqrt{2k})$, which gives
\begin{equation}
    \begin{split}
    \zeta_{\vec{k}} &\sim \frac{Y^{\frac{1}{2}}}{\sqrt{2k}}\left(J_{\frac{3}{2}}(Y) + i Y_{\frac{3}{2}}(Y)\right) =  \frac{1}{\epsilon(\tau)}\sqrt{\frac{A}{2k^3}} (1+ik\tau)e^{-ik\tau}.
\end{split}
\end{equation}
Here, $A$ is a normalization factor that we will determine through cosmic microwave background (CMB) data fitting, and $1/\epsilon(\tau)$ describes the only deviation from the dS invariant distribution of isolated black holes. This is one of the major differences in the scalar fluctuations between our model and a usual field theory inflation model, as explained in Sec \ref{sec:HSTframework}. In a field theory model the factor $\epsilon^{-1}$ would be replaced by its square root, and the correlation function would be evaluated in the slow-roll background. The power spectrum of scalar perturbations is 
\begin{equation}
    \begin{split}
        \braket{\zeta_{\vec{k}}(\tau)\zeta_{\vec{k}'}(\tau)} &= (2\pi)^3\delta(\mathbf{k}+\mathbf{k}')|\zeta_k(\tau)|^2, \\
        \Delta^2_{\mathcal{R}} 
        \equiv \frac{k^3}{2\pi^2} |\zeta_k(\tau)|^2
        &\overset{h.c.}{\sim} \frac{A}{(2\pi)^2}\frac{1}{\epsilon^2(-1/k)}.
    \end{split}
\end{equation}
where $h.c.$ stands for horizon crossing. At horizon crossing $Y\sim 1$, i.e.
\begin{equation}
    \begin{split}
        k \sim He^a &= -\frac{1}{\tau}.
    \end{split}
\end{equation}

\subsection{Fitting to CMB primordial scalar fluctuation power spectrum}\label{sec:fitting}
\begin{figure}
    \centering
    \includegraphics[scale = 0.9]{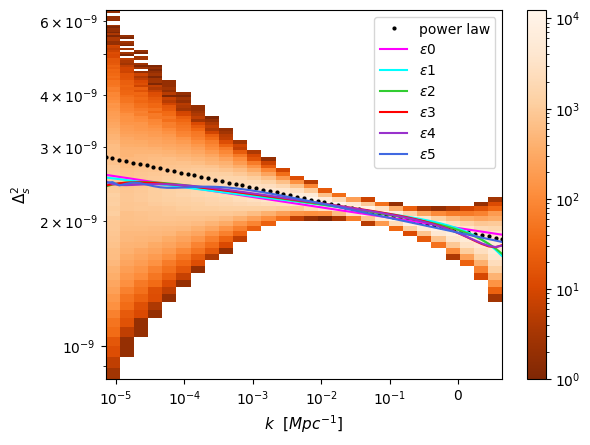}
    \caption{Fitting $\epsilon(\tau)$ with $N=0,\dots,5$ to sample generated using the results from Planck2018 where the scalar tilt $n_s = 0.9649 \pm 0.0042$ at 68\% CL with a running of $d \ln n_s/d\ln k = -0.005 \pm 0.013$ at 95\% CL. The background is a 2D histogram of data sample where the colorbar labels data point density.}
    \label{fig:multifit}
\end{figure}

To determine the exact form of the slow-roll parameter $\epsilon(\tau)$, we fit the analytical scalar power spectrum to the CMB primordial scalar power spectrum by evaluating the slow-roll parameter at the time of horizon crossing for each mode. The $\Lambda$CDM model provides the best current fit to the CMB data. This yields the scalar tilt $n_s = 0.9649 \pm 0.0042$ at 68\% confidence limits, with a running of $d \ln n_s/d\ln k = -0.005 \pm 0.013$ at 95\% confidence limits, according to the Planck 2018 results \cite{Planck}. Expand $\epsilon(-1/k)$ in power series
\begin{equation}
    \begin{split}
        \epsilon(-1/k) &= b_0(k/k_0)^{\alpha} + b_1(k/k_0)^{\alpha+1} + b_2(k/k_0)^{\alpha+2}  + \cdots = \sum_{n=0}^{N} b_n(k/k_0)^{\alpha+n},
    \end{split}
\end{equation}
where $k_0$ in Mpc$^{-1}$ is a scaling factor used to regularize the dimension of wave number $k$, and it can be considered as the CMB pivot scale. Fits up to $N=5$ are shown in Fig. \ref{fig:multifit}. By fitting the function to the data, we observe that for $n\geq 3$, the coefficients $b_n$ in higher-order terms $b_n(k/k_0)^{\alpha+n}$ are significantly smaller than $b_0,b_1,b_2$ and are nearly zero. This is expected due to the power-law fit to the scalar spectrum. Additionally, including these higher power terms does \textit{not} affect the parameters significantly. Thus we can ignore their contribution and only consider the following three cases with $N=0,1,2$. Then, we have
\begin{equation}
    \begin{split}
        \epsilon_N(-1/k) &= \sum_{n=0}^N b_n(k/k_0)^{\alpha+n}    \hspace{5mm} \text{where} \hspace{5mm} N=0,1,2.
    \end{split}
\end{equation}
Data fitting gives $\alpha = (1.23 \pm 0.03)\times 10^{-2}$ when the dominant term is given by $(k/k_0)^{\alpha}$. By scaling conformal time as $\tau \rightarrow k_0\tau$, we obtain $k_0\tau \in [-1,0)$. The dominant terms give the value $\epsilon(\tau) \sim 0.14 \pm 0.01$, as shown in Fig. \ref{fig:eps_distribution}. More discussion about the fitting can be found in Appendix \ref{app:fitting}. 

Written in terms of conformal time we have
\begin{equation}
    \begin{split}
        \epsilon_N(\tau) &= \sum_{n=0}^N b_n\left(-\frac{1}{k_0\tau}\right)^{\alpha+n}, 
    \end{split}
\end{equation}
the reason we do this is to find $\epsilon$ as a function of $\tau$ and to check if $\epsilon \sim \mathcal{O}(0.1)$ at all $\tau$ during inflation. Given that the CMB scalar data reduces to a function of angular momentum, it is not surprising that our model—with a free function $\epsilon(\tau)$—can fit it, as is true for any inflationary model. What truly matters is that the data is highly close to scale invariance. In our case, the fit is more surprising, as the data turns out to be just barely consistent with all our theoretical constraints. Since $\alpha$ is very small, $\epsilon(\tau)$ indeed remains a small number throughout inflation, which can be verified by plugging in the numbers. All the codes are available upon request via email. 
\begin{figure}
        \centering
        \includegraphics[scale = 0.4]{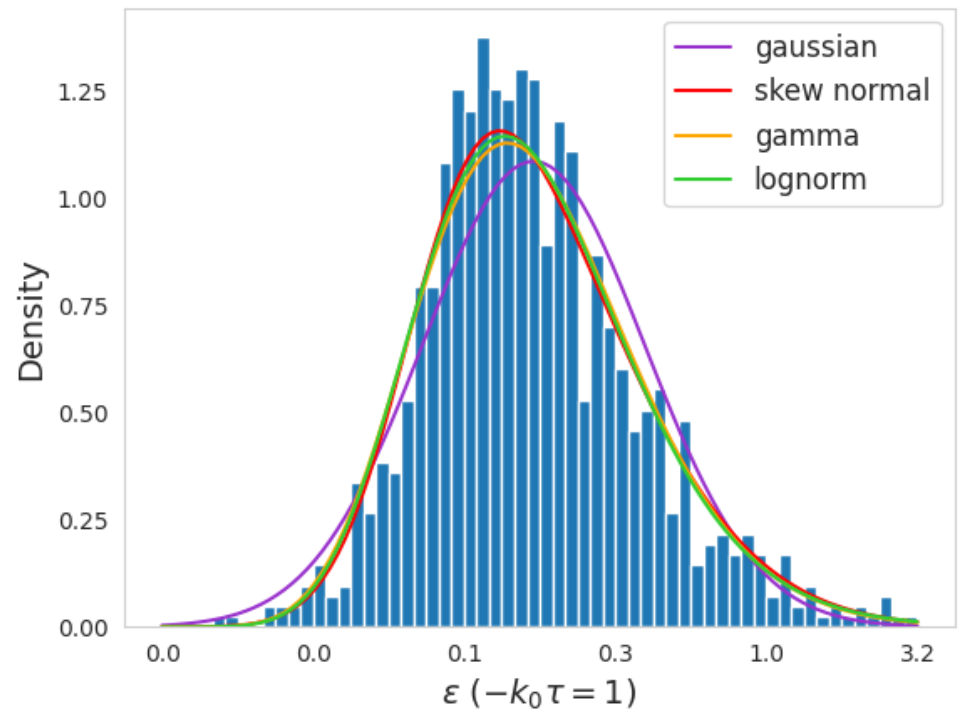}
        \caption{The value of $\epsilon(\tau_i)$ is determined to be $\sim 0.14\pm 0.01$.}
        \label{fig:eps_distribution}  
\end{figure}

\subsection{The number density of the distribution of black holes}
In this section, we determine the Hubble radius $R_I$ at the beginning of inflation in terms of the Planck length $L_P$. To do so, we apply the Carlip-Solodukhin ansatz to the correlator of $\delta H/H$, averaged over a horizon volume $V_{end}$ containing one black hole at the end of slow-roll inflation, as discussed in section \ref{sec:HSTframework},
\begin{equation}\label{eq:csansatz}
    \begin{split}
        \frac{1}{V_{end}}\int_0^{V_{end}} \braket{(\delta H/H)^2} \hspace{1mm} dV &= \frac{n_{BH}}{4\pi(R_I/L_P)^2}, \hspace{10mm} \text{with} \hspace{3mm} n_{BH} = 1.
    \end{split}
\end{equation}
The Carlip-Solodukhin formula $(\Delta S)^2 = S$ for fluctuations gives the factor 
\begin{equation}
    \frac{1}{2\Delta S} = \frac{1}{2\sqrt{\pi(R_I/L_P)^2}}
\end{equation} 
in $\delta H/H$, the black hole entropy is given by the area law $S = \frac{Area}{4G} = \pi(R_I/L_P)^2$, where $L_P$ is the Planck length, and we use $L_P^2 = G = M_P^{-2}$ when $c=\hbar = 1$. In position space, the correlator of $\delta H/H$ is given by the de Sitter invariant Green's function 
\begin{equation}\label{eq:twopint}
    \begin{split}
        \bigg<\frac{\delta H}{H}(\tau_1,\vec{x})\frac{\delta H}{H}(\tau_2,\vec{y})\bigg> &= -\frac{A}{8\pi^2}\ln \left(\frac{-(\tau_1-\tau_2)^2+|\vec{x}-\vec{y}|^2}{4\tau_1\tau_2}\right). 
    \end{split}
\end{equation}
The only difference between the correlator of $\delta H/H$ and the correlator of the scalar fluctuations $\zeta(\tau, \vec{x})$ is a factor of $1/\epsilon^2(\tau)$, which arises from the classical relation between the two in comoving gauge. The normalization factor $A \sim 1.6 \times 10^{-9}$ was determined through data fitting in section \ref{sec:fitting}. By computing the integral and equating it with the Carlip-Solodukhin formula, we get
\begin{equation}\label{eq:csrelation}
    \begin{split}
        \frac{A(1+\ln8)}{12\pi^2} = \frac{n_{BH}}{4\pi(R_I/L_P)^2}.
    \end{split}
\end{equation}
where $n_{BH} = 1$, since at the end of inflation, there is one black hole inside each horizon volume, as discussed in section \ref{sec:HSTframework}. This relation provides an estimation of $R_I$, where
\begin{equation}
    \begin{split}
        R_I = L_P\left(\frac{3\pi}{A(1+\ln 8)}\right)^{\frac{1}{2}} \sim 4.3 \times 10^4L_P.
    \end{split}
\end{equation}
Even though the $R_{end}$ dependency cancels itself in Eq.(\ref{eq:csrelation}), we compute it here as preparation for the angular momentum calculation in section \ref{sec:angmom}. To determine the Hubble radius $R_{end}$ at the end of inflation, we use the power law formula $\epsilon = b(-k_0\tau)^{-\alpha}$ to find the scale factor $e^{a(t)}$. As discussed in section \ref{sec:HSTframework}, slow-roll inflation is required to last long enough such that the wavelengths of order $R_I$ are expanded to the size of the current horizon $1/H_0 \sim  4.4\times 10^3$ Mpc $\sim 8.5 \times 10^{60}L_P$. This leads to 
\begin{equation}\label{eq:expansion}
    \begin{split}
        e^{t_{end}/R_I}\int_0^{t_{end}} e^{-t/R_I} dt &=  \left(e^{t_{end}/R_I}-1\right)R_I \sim 8.5\times 10^{60}L_P \hspace{1mm}  \hspace{5mm}
    \Longrightarrow \hspace{5mm} \frac{t_{end}}{R_I} \sim 129,
    \end{split}
\end{equation}
where $R_I \sim 4.3 \times 10^4 L_P$ is obtained through data fitting. As we can see $\epsilon(\tau) \sim 0.14\pm 0.1$ is almost a constant, we determine the Hubble constant as a function of time by solving Eq.(\ref{eq:ateta1}) and find
\begin{equation}
    H(t) = \frac{1}{0.14  t + R_I} \hspace{10mm} \Longrightarrow \hspace{10mm} 
    R_{end} = \left(129\epsilon  + 1\right)R_I \approx 19.1R_I.
\end{equation}
Note that the Hubble radius is $(e^aH)^{-1}$, but as one can see in Eq.(\ref{eq:csansatz}), the scale factor $e^a$ cancels with its inverse, so we drop $e^{-a}$ in the Hubble radius $e^{-a}R_{end}$. We also comment on the size of $R_{end}$ here. In Eq.(\ref{eq:expansion}) the Hubble constant $H = 1/R_I$ is taken to be almost constant to preserve the de Sitter invariance of the model. As one can see, with a small but non-zero $\epsilon$, $H(t) = R_I^{-1} \rightarrow (\kappa R_I)^{-1}$ where $\kappa \gtrapprox 1$, so it takes longer for wavelength $R_I$ to be expanded to the current horizon size. Therefore, $t_{end} \sim 129\kappa R_I$ which gives a larger $R_{end} >19.1R_I$.

For the consistency of the theory, we also verify that the value of $\epsilon(k)$ stays within the range of $0.1<\epsilon <1$ for very large wave numbers. More specifically, around $k \sim 1 \hspace{1mm} \text{Mpc}/R_I \sim 1.9 \times 10^{52}$, we find
\begin{equation}
    \epsilon(k \sim 1.9\times 10^{52}) = b(k/k_0)^{\alpha} \sim 0.57 < 1,
\end{equation}
where the coefficient $b\sim 0.1$, the pivot scale $k_0 \sim (50 \text{Mpc})^{-1}$, and the power $\alpha \sim 0.012$. the condition $0.1 < \epsilon(k) < 1$ holds for very large $k$ because $\alpha \ll 1$, as we have shown by fitting the analytic scalar spectrum to the CMB data.

\section{Fluctuations in angular momentum and the tensor spectrum normalization}\label{sec:angmomratio}
In momentum space, the tensor part of action is 
\begin{equation}
    \begin{split}
        S_{tensor} &= \int \frac{da d^3k}{8(2\pi)^3} \left(e^{3a}H\gamma_{ij,a}^2 - k^2e^aH^{-1}\gamma_{ij}^2\right) ,\\
        \gamma_{ij}(\tau, \vec{x}) &= \int \frac{d^3k}{(2\pi)^3}\sum_{s=\pm}\epsilon^s_{ij}(k)\gamma_{\vec{k}}^s(\tau)e^{i\vec{k}\cdot\vec{x}},
    \end{split}
\end{equation}
where $\gamma_{ij}^2 = \gamma_{ij}(\vec{k},t)\gamma_{ij}(-\vec{k},t)$ and $M_P^2=L_P^{-2}=1$. The equation of motion is given by the following use $H_a = -\epsilon H$, for each polarization $s$ we have 
\begin{equation}
    \begin{split}
        \gamma^s_{\vec{k},aa} +  \left(3-\epsilon \right)\gamma^s_{\vec{k},a} + k^2H^{-2}e^{-2a}\gamma^s_{\vec{k}}  &= 0.
    \end{split}
\end{equation}
This is the same equation of motion for scalar fluctuations. Following the same argument, we obtain the following expression for tensor modes,
\begin{equation}\label{eq:tensormode}
    \begin{split}
    \gamma^s_{\vec{k}}(\tau) &= \frac{\sqrt{B}}{2\sqrt{k^3}}(1+ik\tau)e^{-ik\tau} ,
\end{split}
\end{equation}
where $B$ is a normalization factor that we will determine later. Note that it differs from the scalar solutions by a factor of $1/\epsilon$, since scalar perturbations and their nearly scale-invariant spectrum arise only in quasi–de Sitter space, with the deviation captured by $\epsilon(\tau)$.

\subsection{Fluctuations in the total angular momentum}\label{sec:angmom}
We use Noether's theorem to find the angular momentum of $\gamma_{ij}$. Consider an infinitesimal rotation that transforms the spatial coordinates as 
\begin{equation}\label{eq:rotationtransform}
    \begin{split}
        x'^k &=  x^k + \frac{1}{2}\xi_{mn}[L^{mn}]^k_l x^l  ,
    \end{split}
\end{equation}
where $\{\xi_{mn}\}$ is a set of infinitesimal parameters, and $L^{mn} = x^m\partial^n-x^n\partial^m$ rotates $x^n$ to $x^m$. Then, the tensor fluctuation fields transform as
\begin{equation}
    \gamma'_{ij}(t,x'^k) = \gamma_{ij}(t,x^k) - \frac{i}{2}\xi_{mn}[S^{mn}]^{pq}_{ij}\gamma_{pq}(t,x^k).
\end{equation}
$S^{mn}$ acts on the tensor fields $\gamma_{ij}$ and follows the tensor representation since $\gamma_{ij}$ has integer spin-2, we have $S^{mn} = L^{mn} \otimes L^{mn}$. $\gamma_{kl}=[S^{mn}]^{ij}_{kl}\gamma_{ij} = ([L^{mn}]^i_k\delta^j_l + [L^{mn}]^j_l\delta^i_k) \gamma_{ij}$, where $[L^{mn}]^i_k$ is the ik-th element in $L^{mn}$ that acts on the first index of $\gamma_{ij}$ and $[L^{mn}]^j_l$ acts on the second index of $\gamma_{ij}$. $[L^{mn}]^i_j = -\delta^i_m\delta^j_n+\delta^i_n\delta^j_m$. We compute the conserved current to find the total angular momentum operator, 
\begin{equation}
    \begin{split}
        [j^{mn}]^{\mu} &= \left[\frac{\partial \mathcal{L}}{\partial (\partial_{\mu}\gamma_{ij})}\partial_{\nu}\gamma_{ij} - \delta^{\mu}_{\nu}\mathcal{L}\right][L^{mn}]^{\nu}_l x^l + \frac{\partial \mathcal{L}}{\partial (\partial_{\mu}\gamma_{ij})}i[S^{mn}]^{pq}_{ij}\gamma_{pq}(t,x^k) \\
        &=  \frac{e^{2a}}{4L_P^2}\gamma'_{ij}\left(L^{mn}\gamma_{ij} + i[S^{mn}]^{pq}_{ij} \gamma_{pq}\right) .
    \end{split}
\end{equation}
Integrating $[j^{mn}]^0$ over one horizon volume at the end of inflation to find the contribution of one local excitations to the total angular momentum $J^{mn}$.
\begin{equation}
    \begin{split}
        J(\tau) &= \int_0^{V_{end}}  j^0 \hspace{1mm} dV = \frac{e^{2a}}{4L_P^2}\int_0^{V_{end}} dV \gamma'_{ij}(\tau, \vec{x})\left(L\gamma_{ij}(\tau, \vec{x}) + i[S]^{pq}_{ij}\gamma_{pq}(\tau, \vec{x})\right) ,
    \end{split}
\end{equation}
where $V_{end}$ is the horizon volume at the end of inflation. Here we restored the Planck length $L_P$. Expanding tensor fluctuations into plane waves. Use Wick's theorem, we find the orbital angular momentum contribution
\begin{equation}
    \begin{split}
        \braket{L^2} &= \frac{e^{4a}}{16L_P^4}\int_0^{V_{end}}d^3y\int_0^{V_{end}}d^3x \braket{\gamma'_{ij,x}L(\vec{x})\gamma_{ij,x}\gamma'_{pq,y}L(\vec{y})\gamma_{pq,y}} \\
        &= \int \left(\braket{\gamma'_{ij,x}\gamma'_{pq,y}}L(\vec{x})L(\vec{y})\braket{\gamma_{ij,x}\gamma_{pq,y}}+L(\vec{y})\braket{\gamma'_{ij,x}\gamma_{pq,y}}L(\vec{x})\braket{\gamma_{ij,x}\gamma'_{pq,y}}\right) ,
    \end{split}
\end{equation}
where $\int = \frac{e^{4a}}{16L_P^4}\int_0^{V_{end}}d^3x\int_0^{V_{end}}d^3y$. Note that, by Wick's theorem, the $\braket{\gamma'_x\gamma_x}$ terms vanish since $\braket{J} = 0$. As shown in the previous section, the solution to the tensor fluctuations differs by the normalization factor, the polarization tensors, and a factor of $4L_P^2$ compared to the solution for $\braket{(\delta H/H)^2}$.
\begin{equation}
    \braket{\gamma^s_{ij}(\tau_1, \vec{x})\gamma^{s'}_{pq}(\tau_2,\vec{y})} = -\delta^{ss'}\epsilon_{ij}\epsilon_{pq}\frac{B}{8\pi^2}\ln\left(\frac{-(\tau_1-\tau_2)^2+|\vec{x}-\vec{y}|^2}{4\tau_1\tau_2}\right).
\end{equation}
The orbital angular momentum has a UV divergence. Let $\Lambda$ be a small distance that sets the UV cutoff, we get
\begin{equation}
    \begin{split}
        \braket{L^2} &= \sum_{s,s',t,t'}\delta^{ss'}\delta^{tt'}\frac{2B^2}{\pi^4}\int \frac{((\vec{x}-\vec{y})\cdot \vec{x})((\vec{x}-\vec{y})\cdot \vec{y})}{|\vec{x}-\vec{y}|^6} \\
        &= \frac{B^2R^4_{end}}{\pi^2L_P^4}\left(\frac{56-165\ln 2}{90} + \frac{1}{2}\ln\left(1+\frac{2R_{end}}{\Lambda}\right)+\frac{R_{end}}{5\Lambda}\right) \\
        &\sim \frac{\lambda B^2R^4_{end}}{\pi^2L_P^4}  \hspace{20mm} \text{for $\lambda \equiv \frac{R_{end}}{\Lambda} \gg 1$}.
    \end{split}
\end{equation} 
The scale factor $e^{a}$ cancels with its inverse in the Hubble radius $e^{-a}R_{end}$. The cross terms $\braket{L(\vec{x})S(\vec{y})}$ and $\braket{S(\vec{x})L(\vec{y})}$ cancel each other. The spin contribution to the total angular momentum is given by,   
\begin{equation}
    \begin{split}
        \braket{S^2} &= \int \braket{\left(\gamma'_{mi,x}\gamma_{ni,x}-\gamma'_{ni,x}\gamma_{mi,x}\right)\left(\gamma'_{mj,y}\gamma_{nj,y}-\gamma'_{nj,y}\gamma_{mj,y}\right) } \\
        &= \int \sum_{s,q,t,r} \braket{\gamma'^s_x\gamma^q_x\gamma'^t_y\gamma^r_y }(\epsilon^s_{mi}\epsilon^q_{ni}\epsilon^t_{mj}\epsilon^r_{nj} - \epsilon^s_{mi}\epsilon^q_{ni}\epsilon^t_{nj}\epsilon^r_{mj} -  \epsilon^s_{ni}\epsilon^q_{mi}\epsilon^t_{mj}\epsilon^r_{nj} + \epsilon^s_{ni}\epsilon^q_{mi}\epsilon^t_{nj}\epsilon^r_{mj}) .
    \end{split}
\end{equation} 
We substitute the general expression for the polarization tensors, computed in appendix \ref{app:polartensor}, and the expression for the tensor modes into $\braket{S^2}$, 
\begin{equation}
    \begin{split}
        \braket{S^2} & = \sum_{s,s',t,t'}\delta^{ss'}\delta^{tt'}\frac{B^2}{4\pi^4}\int \left(-\frac{1}{\tau^2} + \frac{2}{|\vec{x}-\vec{y}|^2}\ln \left(\frac{|\vec{x}-\vec{y}|^2}{4\tau^2}\right)\right) \\
        &=  -\frac{B^2R_{end}^4}{9\pi^2L_P^4} - \frac{7B^2R_{end}^4}{4\pi^2L_P^4} = -\frac{67B^2R_{end}^4}{36\pi^2L_P^2},
    \end{split}
\end{equation}
The UV divergences in $\braket{S^2}$ cancel each other out, and we obtain a finite result without needing a cutoff. More details of the calculations can be found in Appendix \ref{app:angmom}. This gives the total angular momentum fluctuations,
\begin{equation}\label{eq:tensorangmom}
    \begin{split}
        \braket{J^2} = \braket{L^2} - \braket{S^2} &\sim  \frac{\lambda B^2R^4_{end}}{\pi^2L_P^4} .
    \end{split}
\end{equation}
where $\lambda \gg 1$ sets the UV cutoff.

\subsection{Determine the tensor spectrum normalization}
Now, we determine fluctuations in $J$ from Kerr black holes in de Sitter space. The mass of a Kerr black hole can be expressed in terms of its entropy/area and angular momentum, giving us
\begin{equation}
    \begin{split}
        r_+ &= GM + \sqrt{G^2M^2 - a^2} = M\left(1-\frac{a^2}{M^2}\right) \hspace{10mm} \text{where} \hspace{5mm} a = \frac{J}{M},\\
        Area &= 4\pi (r_+^2 + a^2) = 8\pi M^2 \bigg( 2 - \frac{J^2}{2M^4} + \mathcal{O}\big(J^4\big) \bigg).
    \end{split}
\end{equation}

The entropy as a distribution of $J$ with fixed $M$ (where M fluctuates as well, but including both $\Delta J$ and $\Delta M$ is a second-order effect and thus negligible) is given by
\begin{equation}
    \begin{split}
        S = \frac{Area}{4G} = \frac{4\pi M^2}{G} -\frac{\pi J^2}{GM^2}. 
    \end{split}
\end{equation}
The von Neumann entropy is $S= -tr (\rho \ln \rho)$ where $\rho = e^{-K}$ is the density matrix (which must be invariant under dS group, in terms of 2-point function computed from density matrix, the probability distribution for each geodesic must be invariant under the dS group) and $K$ is modular Hamiltonian, and we have $tr\rho =1$. The number of states is given by
\begin{equation}
    \begin{split}
        \Omega \sim e^{S} = e^{\frac{4\pi M^2}{G}}e^{-\frac{\pi J^2}{GM^2}},
    \end{split}
\end{equation}
where we dropped a normalization factor. Since we have $\braket{J}=0$, the fluctuation in $J$ for each rotating black hole is 
\begin{equation}\label{eq:kerr}
    \braket{J^2} - \braket{J}^2 = \braket{J^2} = \frac{GM^2}{2\pi} = \frac{R_I^2}{8\pi L_P^2}.
\end{equation}
The black holes have a radius $R_I \sim 4.3 \times 10^4L_P$, which is just the Schwarzschild radius $R_I = 2GM$ because they have $\braket{J}=0$. Since these black holes, i.e. the isolated excitations, provide a source of tensor fluctuations, we equate Eq.(\ref{eq:tensorangmom}) and Eq.(\ref{eq:kerr}), and find
\begin{equation}
    \begin{split}
        B &= \sqrt{\frac{\pi}{8\lambda}}\frac{R_I^2}{R^2_{end}}\frac{L_P}{R_I} \sim  \frac{4\times 10^{-8}}{\sqrt{\lambda}}.
    \end{split}
\end{equation}
Then the tensor-to-scalar ratio $r$ is found to be
\begin{equation}
    \begin{split}
        r = \frac{B}{A/\epsilon^2} \sim \frac{1}{2\sqrt{\lambda}},
    \end{split}
\end{equation}
we see that when $\lambda > 2\times 10^2$, i.e. with a UV cutoff $\Lambda \sim 10^3 L_P$, it gives a ratio that satisfies the observations, where the constraint on the tensor-to-scalar ratio \cite{BICEP:2021xfz} is $r<0.036$ at $95\%$ confidence limit. In other words, at the end of inflation, the maximum fraction of the horizon size needs to be cut off is about $\mathcal{O}(10^{-3})$ (recall that $t_{end} \sim 10^6L_P$) for the tensor-to-scalar ratio to match with observations, and the ratio gets smaller if a smaller fraction is removed. 

\section{Conclusions}
In this paper, we first determine the constant normalization factor $A$ and the slow-roll parameter $\epsilon(\tau)$ in our model through fitting the scalar power spectrum to the CMB data.Using the result of $A$ in the Carlip-Solodukhin entropy formula, we relate the Hubble radius $R_I$ at the beginning of inflation to the Planck length $L_P$. The slow-roll parameter $\epsilon$ provides a relationship between the Hubble radius at the beginning of inflation and end of inflation, with $R_{end} \sim \mathcal{O}(10^2)\epsilon R_I$. Then we obtain the normalization factor $B$ in the tensor power spectrum by computing angular momentum fluctuations for a de Sitter invariant distribution of black holes, and we find the tensor spectrum normalization. Finally, the tensor-to-scalar ratio depends on the choice of a UV cutoff and $\epsilon$, and the ratio satisfies the observational constraints, where $r < 0.036$. 

It is not surprising that our model can fit the CMB data.  Like conventional field theory inflation, we have one free function, the slow-roll scale factor $a(t)$, which has to fit one function of angular momentum over a few decades in $L$.  The relation between $a(t)$ and the data is different in the two classes of models, but both incorporate a suppression of tensor version scalar modes and non-Gaussian versus Gaussian fluctuations, which follows from the general principles of classical cosmological perturbation theory \cite{maldafluct}.  

It is when we look at the detailed theoretical assumptions underlying our model that we see how easily our fit could have failed.  The overall amplitude of scalar fluctuations is of order
$\epsilon^{-1} (L_P/R_I)$ .   In HST models $\epsilon$ is constrained to lie in a range $1 > \epsilon > 0.1 $ (unless the constant $C$ in the fast scrambling rate is significantly different than $1$), so $R_I$ cannot be larger than about $10^6 L_P$, which implies many e-folds of inflation.  If the data had not shown that $1 - n_s \ll 1$ our model would have been theoretically inconsistent.    Similarly, we are consistent with the observed absence of tensor fluctuations, despite our theoretical constraint on $\epsilon$, only because the model predicts a different $\epsilon$ dependence for the tensor to scalar ratio than field theory models.  

We should note however that there is some theoretical wiggle room in HST models.  As far as we are aware, there is no calculation in the literature that determines the constant $C$ in the fast scrambling time from the Einstein equations.  We are interested in a particular scrambling time, the time it takes to equilibrate a certain set of constrained q-bits.   In addition, any gravity based calculation would only be the first term in an asymptotic expansion in the Planck length over the diamond size and we need the result for a diamond the size of the microscopic inflationary horizon.  There is no principle that says that $C$ cannot be a function of this ratio.  Indeed, from the point of view of the microscopic models of \cite{hilbertbundles} fast scrambling is caused by a Thirring coupling between the fermion fields and that coupling could be allowed to vary with the diamond size without violating any principles of which we're aware. 

The HST model makes different predictions for non-Gaussian fluctuations than field theory models, but since these have to do with three-point functions of the tensor modes, they cannot be tested at present, as fluctuations in the tensor modes have not yet been detected. If primordial non-Gaussianity is observed in the future, the predictions may become testable through comparison with observational data.  More significant is that its post-inflationary history as a dilute gas of black holes gives rise to an early era of structure formation, a possible origin for the baryon asymmetry, and, with the additional hypothesis of a discrete gauge symmetry, a model for dark matter as Planck mass black hole remnants with discrete gauge charge.  

We also want to reiterate the conceptual differences between the arguments for approximate homogeneity, isotropy and scale invariance of the two-point functions in field theory models and HST models of inflation.  In field theory models it is hypothesized that inhomogeneities in initial field configurations have been ``inflated away" by the expansion of the universe.  Essentially by definition, this introduces the ``trans-Planckian problem": the modes we observe in the sky originated as modes whose wavelength was shorter than the Planck length.  The only argument ever produced to get around this was based on the adiabatic theorem: that as those modes are brought into the range where effective field theory is valid they are adiabatically brought into the ``Bunch Davies vacuum" state.  It has never been clear how this applies to a set up in which there are simply no modes with wavelength shorter than the Planck scale.  

On a more fundamental level, this argument ignores the fact that during slow-roll inflation there is actually no physical expansion of the horizon.  That is, at the FRW time corresponding to the end of inflation, the causal diamond of a detector is much smaller than the causal diamond that will eventually be seen on that surface in the far future.  So the inflationary stretching of modes is something that local detectors cannot measure.  This is well known for dS space in static coordinates, where the effect of inflation is merely to make anything not tied to the detector fade into the equilibrium state on the cosmic horizon.  It's only locally near the horizon that the stretching of modes remains detectable at late static time.  By combining this local stretching symmetry around each black hole horizon, with a theoretically motivated translation and rotation symmetry of the probability distribution of black holes {\it in flat FRW coordinates}, we get the usual dS invariant formula for the two-point function of $\frac{\delta H}{H}$ in flat coordinates. 

As we've noted, in HST models, homogeneity and isotropy follow from the $S_N$ gauge symmetry of our block diagonal matrix initial condition on the holographic screen.  One may question why one has to take equal size blocks.  The answer is environmental selection. From the point of view of bulk physics, unequal blocks would correspond to a gas of black holes of very different masses. This would lead to rapid collisions and agglomeration into large stable black holes and no Hot Big Bang.  Nothing like our present universe could ever evolve.  It's of course possible to postulate slightly unequal size blocks and break the $S_N$ symmetry.  This would mean some extra small inhomogeneity apart from that which follows from general principles.  While we cannot rule that out, it seems natural to make predictions based on the minimal, unavoidable, inhomogeneities.

\begin{center}
\textbf{Acknowledgements }
\end{center} 

The work of T. B. and S. A is supported in part by the DOE under grant DE-SC0010008.  We would like to thank W. Fischler for many clarifying remarks on the manuscript and N. Fernandez for help with understanding the CMB codes. 

\appendix 
\section{Matrix Models, Black Holes and de Sitter Space}\label{app:review}

Consider an $N \times (N + 1)$ matrix $\psi_i^A$ of two dimensional Dirac fields, living on an interval.  We also impose a UV cutoff on spatial momenta, which is large enough that Cardy's formula is approximately valid for each individual field.  We choose a Hamiltonian $K(N)$ that is $1/N$ times a trace of a quadratic polynomial in $M_i^j = N^{-1} \bar{\psi}_i^A \Gamma \psi_A^j$ .  By 't Hooft scaling, the interaction time scale for this Hamiltonian is $\sim N$.  The interactions are invariant under the ``fuzzy" approximation to area preserving diffeomorphisms of the two sphere, and they can also be chosen to be approximately invariant under $1 + 1$ dimensional conformal transformations.  $K(N)$ is the modular Hamiltonian of a causal diamond of entropy $\sim N^2$ and 
\begin{equation} 
    e^{iK(N)} e^{- i K(N + 1)} . 
\end{equation} 
is the unitary embedding that acts as a time evolution operator between proper times $NL_p$ and $(N + 1) L_P$, with time measured from the Big Bang along some geodesic in a flat FRW space-time.  

We now state without detailed proof some features of this model.  Viewed as an ever expanding cosmology we can calculate the total entropy and energy from the model and then use geometry to calculate the energy and entropy densities as a function of time.  We find 
$\sigma \sim \sqrt{\rho} \sim t^{-1}$, the Friedman equation for a $p = \rho$ universe.  If we stop the expansion of the Hilbert space at some finite $N$ and continue to evolve with $K(N)$ we obtain the Carlip-Solodukhin model of a static horizon. 

Viewing this as the equilibrium state of empty de Sitter space, we can define constrained states of that system in which one forces a small block of the matrix $M_i^j$ to be temporarily decoupled from the rest of the fields, by forcing the off diagonal fields to be zero.   This situation will persist for a time of order $N {\rm ln}\ N$.  The probability for this to be found accidentally in the equilibrium density matrix has a thermal form with a temperature proportional to $N^{-1}$ if we consider the size of the small block to be its energy.  This is consistent with the idea that the small block is a black hole, because it is an equilibrium system, with entropy that scales like the square of its energy, whose modular Hamiltonian satisfies the Carlip-Solodukhin relationship.  

Considering states with two small blocks, we can even calculate the Newton force between them.  It arises in second order perturbation theory by virtually turning on and off the off diagonal blocks connecting each of them to the large set of horizon degrees of freedom.  The instantaneous distance between them comes into the calculation because they cannot interact until they enter into the same causal diamond.  The ``action at a distance" force is generated by purely causal scattering processes involving degrees of freedom spread over diamond boundaries.  

\section{Fitting the scalar power spectrum to the CMB data}\label{app:fitting}
\begin{figure}[h]
    \begin{minipage}{0.33\textwidth}
        \centering
        \includegraphics[width=1\linewidth]{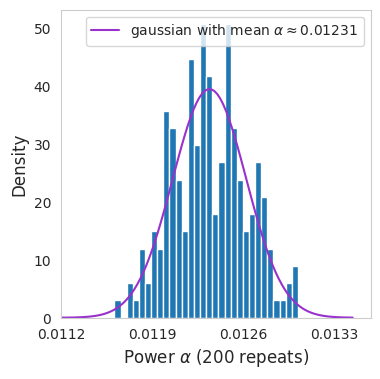}
    \end{minipage}
    \begin{minipage}{0.33\textwidth}
        \centering
        \includegraphics[width=1\linewidth]{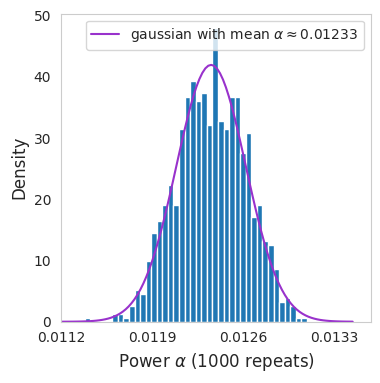}
    \end{minipage}
    \begin{minipage}{0.33\textwidth}
        \centering
        \includegraphics[width=1\linewidth]{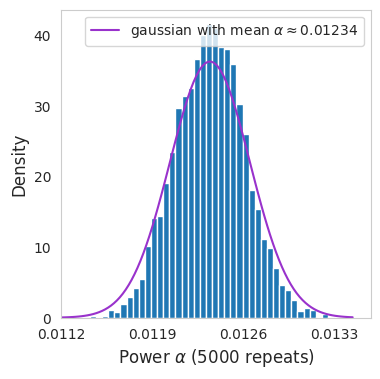}
    \end{minipage}
    \caption{Repeating $curve\_fit$ 200, 1000 and 5000 times to determine the parameter $\alpha$, we find $\alpha = 0.012\pm 0.001$. The results are all within one sigma region compared to each other.}
    \label{fig:diffrepeats}
\end{figure}

We have four parameters in the analytical scalar power spectrum, the normalization factor $A$, the coefficient $b$, the power $\alpha$ of the wave number, and the pivot scale $k_0$ in $\epsilon(k) = b(k/k_0)^{\alpha}$. These parameters are determined by repeating the $curve\_fit$ function in Python. We obtained results from 200, 1000 and 5000 repeats, and the results show no significant differences, an example for parameter $\alpha$ is shown in Fig. \ref{fig:diffrepeats}. Therefore, we use the results from 1000 repeats in this paper.  

\begin{figure}[h]
    \begin{minipage}{0.5\textwidth}
        \centering
        \includegraphics[width=1\linewidth]{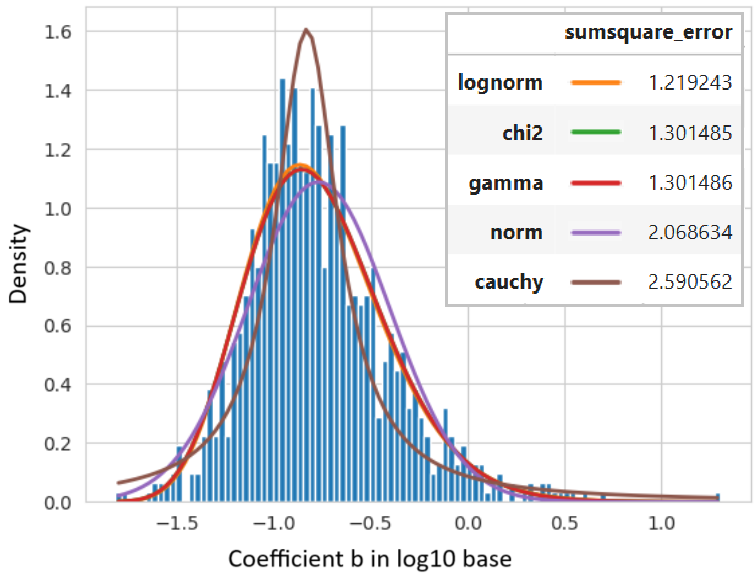}
    \end{minipage}
    \begin{minipage}{0.5\textwidth}
        \centering
        \includegraphics[width=1\linewidth]{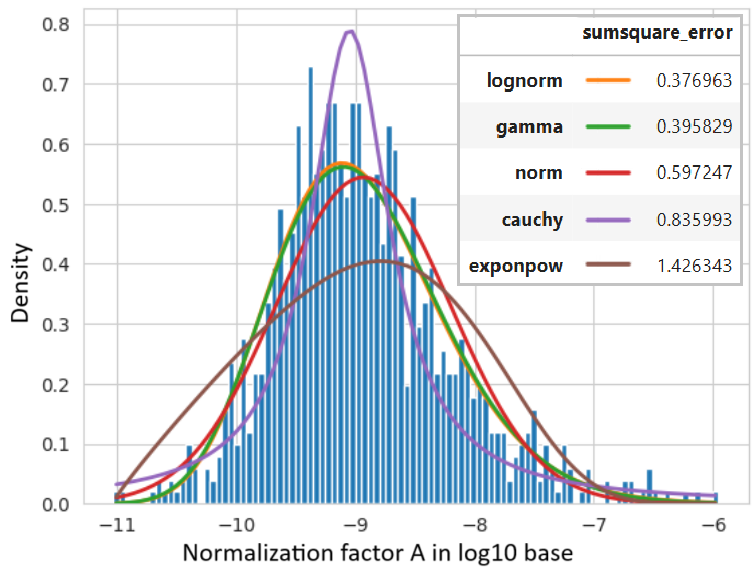}
    \end{minipage}
    \caption{Determine the best fit using the least sum of squared residuals method, we find the distributions of the coefficient $b$ and the normalization factor $A$ follow ``nearly normal" log-normal curves. This skewed behavior is due to the fact that the amplitude is a product $Ab^{-2}$ which leads to an uncertainty.}
    \label{fig:bestfit}
\end{figure}

We determine the best-fitting distribution of each set of data for each parameter using the $Fitter$ library in Python \cite{fitter}, which compares the sum of squared residuals method among common distributions. The candidates are normal, log-normal, Chi-squared, Gamma, Cauchy and exponential power distributions as shown in the legends. After taking the logarithmic scale, the distributions of the power $\alpha$ and the pivot scale $k_0$ are best described by the normal distribution. In the log scale, the normalization factor $A$ and the coefficient $b$ follow log-normal curves, shown in Fig.\ref{fig:bestfit}, indicating a gentle ``exponential of exponential" behavior in the data distribution. This comes from the fact that the overall normalization of the spectrum is given by the product of $A$ and $b$, i.e. $Ab^{-2}$, this multiplicative nature in the amplitude leads to an uncertainty and introduces log-normal behavior in the distribution of fitted parameters. However, the peaks of each log-normal distribution lie within one sigma region of the corresponding normal distribution, and the order of magnitude of our results remains unaffected by choosing either log-normal or normal distributions.  

\section{Compute the polarization tensors}\label{app:polartensor}
The polarization tensors $\epsilon_{ij}^s(k)$ are transverse $k^i\epsilon_{ij}^s(k)=0$, traceless $\eta^{ij}\epsilon_{ij}^s(k)=0$ and purely spatial $\epsilon_{0\mu}=0$, conjugate $\epsilon^{s*}_{ij}(k) = \epsilon^s_{ij}(-k)$ by definition, and $\epsilon^s_{ij}\epsilon^{s'}_{ij}=2\delta^{ss'}$. Where $\epsilon^+_{ij} = \hat{u}_i\hat{u}_j - \hat{v}_i\hat{v}_j $ and $\epsilon^-_{ij} = \hat{u}_i\hat{v}_j+\hat{v}_i\hat{u}_j $. Where $\hat{u}, \hat{v}$ are unit vectors orthogonal to $\hat{n}$ and orthogonal to each other. When $\hat{n}=\vec{k}/k$ in $z$ direction, and $\hat{u}=\hat{x}$ and $\hat{v}=\hat{y}$. In order to find polarization tensor $\epsilon^s_{ij}(k)$, write down $\hat{u}, \hat{v}$ in spherical coordinates for $\vec{k}$ general direction, it is equivalent to coordinate transformation by rotating $\hat{z}$ to $\vec{k}$ direction, then $\hat{u} = \hat{x}'$ and $\hat{v} = \hat{y}'$, we have
\begin{equation}
    \begin{split}
        n^1 = \sin\theta \cos\varphi, \hspace{10mm} n^2 = \sin\theta \sin\varphi, \hspace{10mm} n^3 = \cos\theta,
    \end{split}
\end{equation}
where $\theta$ is the polar angle and $\phi$ is the azimuthal angle in spherical coordinates, the system $\hat{n}, \hat{u}, \hat{v}$ can be considered as first rotate $xy$-plane about $\hat{z}$ by $\varphi$, then rotate $xz$-plane about $\hat{u}$ by $\theta$. Use Euler angle $z-y'-x''$ intrinsic rotation (or $x-y-z$ extrinsic rotation)
\begin{equation}
    \begin{split}
        R_x(\phi) = \begin{pmatrix}
            1 & 0 & 0 \\
            0 & \cos\phi & -\sin\phi \\
            0 & \sin\phi & \cos\phi
        \end{pmatrix}&,\hspace{20mm}
        R_y(\theta) = \begin{pmatrix}
            \cos\theta & 0 & \sin\theta \\
            0 & 1 & 0 \\
            -\sin\theta & 0 & \cos\theta
        \end{pmatrix}, \\
        R_z(\varphi) = &\begin{pmatrix}
            \cos\varphi & -\sin\varphi & 0 \\
            \sin\varphi & \cos\varphi & 0 \\
            0 & 0 & 1
        \end{pmatrix},
    \end{split} 
\end{equation}
\begin{equation}
    \begin{split}
        R = R_z(\varphi)R_y(\theta)R_x(\phi=0) = \begin{pmatrix}
            \cos\theta\cos\varphi & -\sin\varphi & \sin\theta \cos\varphi \\
            \cos\theta \sin\varphi & \cos\varphi & \sin\theta \sin\varphi \\
            -\sin\theta & 0 & \cos\theta
        \end{pmatrix}.
    \end{split}
\end{equation}
Then we have $\hat{n}=R\hat{z}, \hat{u}=R\hat{x}, \hat{v}=R\hat{y}$
\begin{equation}
    \begin{split}
        \hat{n} = \begin{pmatrix}
            \sin\theta \cos\varphi \\
            \sin\theta \sin\varphi \\
            \cos\theta
        \end{pmatrix}, \hspace{15mm} \hat{u} = \begin{pmatrix}
            \cos\theta\cos\varphi \\
            \cos\theta\sin\varphi \\
            -\sin\theta
        \end{pmatrix}, \hspace{15mm} \hat{v} = \begin{pmatrix}
            -\sin\varphi \\
            \cos\varphi \\
            0
        \end{pmatrix}.
    \end{split}
\end{equation}
Now write down the most general expression for $\epsilon_{ij}(k)$
\begin{equation}
    \begin{split}
        \epsilon_{ij}^+ &= \left(
\begin{array}{ccc}
 \cos ^2\theta \cos ^2\varphi-\sin ^2\varphi & \left(\cos ^2\theta+1\right) \sin \varphi \cos \varphi & -\sin \theta \cos
   \theta \cos \varphi \\
 \left(\cos ^2\theta+1\right) \sin \varphi \cos \varphi & \cos ^2\theta \sin ^2\varphi-\cos ^2\varphi & -\sin \theta \cos
   \theta \sin \varphi \\
 -\sin \theta \cos \theta \cos \varphi & -\sin \theta \cos \theta \sin \varphi & \sin ^2\theta \\
\end{array}
\right), \\
        \epsilon^-_{ij} &= \left(
\begin{array}{ccc}
 -2 \cos \theta \sin \varphi \cos \varphi & \cos \theta \cos (2 \varphi ) & \sin \theta \sin \varphi \\
 \cos \theta \cos (2 \varphi ) & \cos \theta \sin (2 \varphi ) & -\sin \theta \cos \varphi \\
 \sin \theta \sin \varphi & -\sin \theta \cos \varphi & 0 \\
\end{array}
\right).
    \end{split}
\end{equation}

\section{Fluctuations in the total angular momentum}\label{app:angmom}
In this appendix we write down ingredients that we need to compute the total angular momentum, for the tensor fluctuations, the two-point Green's function in position space is given by Eq.(\ref{eq:twopint}) with a different normalization constant $B$, 
\begin{equation}
    \begin{split}
        \braket{\gamma(\tau, \vec{x})\gamma(\tau, \vec{y})} &= -\frac{B}{8\pi^2}\ln\left(\frac{|\vec{x}-\vec{y}|^2}{4\tau^2}\right), \\
        \braket{\gamma'(\tau, \vec{x})\gamma(\tau,\vec{y})} &= \frac{B}{8\pi^2\tau} = \braket{\gamma(\tau, \vec{x})\gamma'(\tau,\vec{y})},  \\ 
        \braket{\gamma'(\tau, \vec{x})\gamma'(\tau,\vec{y})}&= -\frac{B}{4\pi^2}\frac{1}{|\vec{x}-\vec{y}|^2}, 
    \end{split}
\end{equation}
\begin{equation}
    \begin{split}
        L(\vec{y})\braket{\gamma'(\tau, \vec{x})\gamma(\tau,\vec{y})} & = 0
        = L(\vec{x})\braket{\gamma(\tau, \vec{x})\gamma'(\tau,\vec{y})}
    \end{split}
\end{equation}
\begin{equation}
    \begin{split}
        L(\vec{x})L(\vec{y})\braket{\gamma(\tau, \vec{x})\gamma(\tau, \vec{y})} &= \frac{B}{8\pi^2}(\vec{x}\times \vec{\nabla}_x)\cdot (\vec{y}\times \vec{\nabla}_y)\ln\left(\frac{|\vec{x}-\vec{y}|^2}{4\tau^2}\right) \\
        &= -\frac{B}{2\pi^2}\frac{(|\vec{x}-\vec{y}|\cdot \vec{x})(|\vec{x}-\vec{y}|\cdot \vec{y})}{|\vec{x}-\vec{y}|^4}
    \end{split}
\end{equation}
\begin{equation}
    \begin{split}
    \braket{\gamma'_x\gamma'_y}L(\vec{x})L(\vec{y})\braket{\gamma_x\gamma_y} &= \frac{B^2}{8\pi^4}\frac{(|\vec{x}-\vec{y}|\cdot \vec{x})(|\vec{x}-\vec{y}|\cdot \vec{y})}{|\vec{x}-\vec{y}|^6} \\
    L(\vec{y})\braket{\gamma'_x\gamma_y}L(\vec{x})\braket{\gamma_x\gamma'_y} &= 0 
    \end{split}
\end{equation}
\begin{equation}
    \begin{split}
        \braket{\gamma'(\tau, \vec{x})\gamma(\tau, \vec{y})}\braket{\gamma(\tau, \vec{x})\gamma'(\tau, \vec{y})} &= \frac{B^2}{64\pi^4\tau^2}, \\
        \braket{\gamma'(\tau, \vec{x})\gamma'(\tau,\vec{y})}\braket{\gamma(\tau,\vec{x})\gamma(\tau, \vec{y})} &= \frac{B^2}{32\pi^4}\frac{1}{|\vec{x}-\vec{y}|^2}\ln\left(\frac{|\vec{x}-\vec{y}|^2}{4\tau^2}\right).
    \end{split}
\end{equation}
We also have a factor of 16 from summing over the polarization tensors in appendix \ref{app:polartensor} and a factor of 4 from summing over polarization $\sum_{ss'}\delta^{ss'}=2$. The integration over the horizon volume at the end of inflation is straightforward to do, and using the above formulae one can check that the UV divergences in $\braket{S^2}$ appear in the form of $\ln((x-y)^2)$ when $|\vec{x}-\vec{y}|\rightarrow 0$ and $\ln((x+y)^2)$ when $\vec{x}, \vec{y} \rightarrow 0$, the coefficients of the positive and negative terms cancel precisely cancel each other. A cutoff is needed for the UV divergence in $\braket{L^2}$.

\end{document}